\documentclass[useAMS,usenatbib]{mnras}
\usepackage{graphicx,amssymb,amsmath,times,verbatim,pdfpages,multirow,booktabs,array,dcolumn}
\usepackage[caption=false]{subfig}
\usepackage[flushleft]{threeparttable}

\title[Variability of PG 1553+113]{Multi-band Variability of the TeV Blazar PG 1553+113 with {\it XMM-Newton}}
\author[Dhiman et al.]{Vinit Dhiman$^{1,2}$\thanks{Email: dhiman@aries.res.in}, Alok C. Gupta$^{1}$\thanks{Email: acgupta30@gmail.com}, Haritma Gaur$^{1}$\thanks{Email: harry.gaur31@gmail.com}, Paul J. Wiita$^{3}$ \\
\\
$^{1}$Aryabhatta Research Institute of Observational Sciences (ARIES), Manora Peak, Nainital 263001, India \\
$^{2}$School of Studies in Physics \& Astrophysics, Pt.\ Ravishankar Shukla University, Amanaka G.E. Road, Raipur 492010, India \\
$^{3}$Department of Physics, The College of New Jersey, PO Box 7718, Ewing, NJ 08628-0718, USA}

\begin{document}
\label{firstpage}
\pagerange{\pageref{firstpage}--\pageref{lastpage}}
\maketitle

\begin{abstract}
\noindent
We present variability analyses of twenty pointed {\it XMM-Newton} observations of the high energy peaked TeV blazar PG 1553+113 taken during 2010 to 2018. We found intraday variability in the total X-ray energy range (0.3 -- 10 keV) in 16 out of 19 light curves or a duty cycle of  $\sim$ 84\%. A discrete correlation function analysis of the intraday light curves in the soft and hard X-ray bands peaks on zero lag, showing that the emission in hard and soft bands are co-spatial and emitted from the same population of leptons. Red-noise dominates the power spectral density (PSD) of all the LCs although the PSDs have a range of spectral slopes  from $-$2.36 to $-$0.14. On longer timescales, the optical and UV variability patterns look almost identical and well correlated, as are the soft and hard X-ray bands, but the optical/UV variations are not correlated to those in the X-ray band, indicating that  the optical/UV and X-ray emissions are emitted by two different populations of leptons. We briefly discuss  physical mechanisms which may be capable of explaining the observed flux and spectral
variability of PG 1553+113 on these diverse timescales.
\end{abstract}


\section{Introduction} \label{sec:introdues}

\noindent
Radio-loud active galactic nuclei (AGN) eject relativistic charged particle jets from their central supermassive black holes (SMBHs) which are in the mass range 10$^{6} -$ 10$^{10} M_{\odot}$. Blazars are a subclass thereof, where one of the two-sided jets approaches the observer within an angle of $\lesssim$ 10$^{\circ}$ to the line of sight \citep[e.g.,][]{1995PASP..107..803U}. Blazar jets are capable of transporting large amounts of power in the forms of kinetic energy, radiation, and magnetic fields, from the vicinity of the central SMBH.   Blazars are historically considered to be the combination of the BL Lacertae (BL Lac) objects, which show featureless spectra or very weak emission lines  \citep[equivalent width EW $\leq$ 5\AA;][]{1991ApJ...374...72S,1996MNRAS.281..425M}, and flat spectrum radio quasars (FSRQs) with prominent emission lines in their composite optical/UV spectrum \citep{1978PhyS...17..265B,1997A&A...327...61G}. The name ``blazar'' indicates that they show flux and often spectral variability across the entire electromagnetic (EM) spectrum.  They are also characterised by core dominated radio structures,  emit predominantly non-thermal radiation, and when measured, they evince strong polarisation in not only the radio but also optical bands (where $\geq$ 3\% qualifies as strong). \\
\\
Since blazars emit radiation throughout the EM spectrum, simultaneous (or nearly so) multi-wavelength (MW) observations produce their spectral energy distributions (SEDs) in the energy range from radio through $\gamma-$rays, to at least GeV energies and sometimes to the very high TeV energies. The MW SEDs of blazars in a log($\nu F_{\nu}$) vs. log($\nu$) representation show double-humped structures \citep{1998MNRAS.299..433F} which are used to classify them as high synchrotron peaked (HSP), intermediate synchrotron peaked (ISP), or low synchrotron peaked (LSP) blazars. The first (low-energy) hump lies in IR to optical bands in LSPs (also called LBLs, or low frequency/energy BL Lacs) and in the far-ultraviolet to X-ray band in HSPs (high-frequency/energy BL Lac objects, or HBLs). The second (high-energy) hump peaks at GeV energies in LSPs and at TeV energies in HSPs. Based on the synchrotron peak frequency($\nu _{s}$), blazars classified into categories HSPs, ISPs, and LSPs, having $\nu _{s} > 10^{15}$ Hz, $10^{14} Hz < \nu _{s} < 10^{15}$ Hz, and  $\nu _{s} < 10^{14}$ Hz, respectively, \citep[e.g.,][]{2010ApJ...716...30A}. The lower energy hump of the SED is clearly dominated by synchrotron radiation which originates from ultrarelativistic electrons in the jets but the origin of the high energy SED remains in question. According to leptonic models, the high energy portion of the SED can be explained by synchrotron photons (Synchrotron Self Compton, SSC) or external photons (External Compton, EC) gaining energy thorough  inverse Compton (IC) scattering  from the same electrons that emit the synchrotron photons \citep[e.g.,][]{2007Ap&SS.307...69B}. Alternate hadronic models for the high energy SED, in which the high energy photons are produced by the synchrotron emission from relativistic protons and/or the secondary charged pion decay process resulting from  photon-hadronic interactions may be preferable for some sources \citep[e.g.,][]{1993A&A...269...67M,2013ApJ...768...54B}. \\
\\
Flux variations in blazars that are observed to occur on rapid timescales, from a few minutes to less than a day, are commonly known as intraday variability (IDV) \citep{1995ARA&A..33..163W} or microvariability \citep{1989Natur.337..627M} or intranight variability \citep{1993MNRAS.262..963G}. Variations in flux between days and a few weeks is often called short term variability (STV), while the changes in flux observed over the course of  a few months to several years are generally known as long term variability (LTV) \citep{2004A&A...422..505G}. \\
\\
PG 1553+113 ($\alpha_{\rm 2000}$ = 15h 55m 43.04s; $\delta_{\rm 2000} = +11^{\circ} 11^{\prime} 24.36^{\prime\prime}$) at $z = 0.433$ 
\citep{2010ApJ...720..976D,2019ApJ...884L..31J} was first detected in the Palomar-Green (PG) survey of UV-excess stellar sources as a 15.5 mag blue stellar object \citep{1986ApJS...61..305G}, and classified as BL Lac object by its featureless spectrum with an optical R magnitude varying from $\sim$13 to $\sim$15.5 mag \citep{1983BAAS...15..957M}. It is sometimes classified as an extreme HBL, based on the synchrotron peak position in its SED \citep{1990PASP..102.1120F}, and flux ratio of 5 GHz radio flux $F_{\rm 5GHz}$ and 2 keV X-ray flux $F_{\rm 2keV}$ which should be log ($F_{\rm 2keV} / F_{\rm 5GHz}$) $\geq -4.5$ for that subclass \citep{2003AJ....125.1060R}. Values of log  ($F_{\rm 2keV} / F_{\rm 5GHz}$) from  $-$4.99 to $-$3.88 have been measured for PG 1553+113 \citep{2003AJ....125.1060R,2006AJ....132..873O}. \\
\\
 VHE $\gamma-$ray emission above 200 GeV from PG 1553+113 was discovered using observations from the {\it High Energy Stereoscopic System} (H.E.S.S.) \citep{2006A&A...448L..19A}. This blazar has been observed on multiple other occasions with both the space-based {\it Fermi} and ground-based $\gamma-$ray telescopes e.g. {\it H.E.S.S., MAGIC} (Major Atmospheric Gamma Imaging Cherenkov Telescopes), {\it VERITAS}  (Very Energetic Radiation Imaging Telescope Array System), etc. \citep[e.g.,][and references therein]{2007ApJ...654L.119A,2009A&A...493..467A,2008A&A...477..481A,2010ApJ...708.1310A,2012ApJ...748...46A,2015ApJ...799....7A}. Its $\gamma-$ray emission is variable, which is typical for TeV blazars  \citep{2015ApJ...802...65A,2015MNRAS.450.4399A}. \\
\\
PG 1553+113 is a bright X-ray blazar as well and has been observed by several X-ray telescopes across large energy ranges, e.g., {\it Einstein, ROSAT, ASCA, BeppoSAX, RXTE, MAXI, XMM-Newton, Chandra, Swift, Suzaku}, and {\it NuSTAR} over many epochs in different flux states, but for most of these observations no evidence of rapid flux variability was noticed on IDV timescales \citep[e.g.,][and references within]{2008ApJ...682..775R}. However, one X-ray observation with {\it XMM-Newton}  (out of 6 examined) showed rapid IDV on a time scale of around one hour \citep{2017MNRAS.466.3762R}. A  Whole Earth Blazar Telescope (WEBT) multiwavelength campaign for the blazar PG 1553+113 was coordinated  in 2013 during which the X-ray spectrum remained stable and a bluer-when-brighter trend in the optical region was found \citep{2015MNRAS.454..353R}. 
An {\it RXTE} observational campaign found a doubling of the X-ray flux  in 10 days and spectral curvature  \citep{2006AJ....132..873O}. In a {\it Suzaku} observation, PG 1553+113 exhibited spectral curvature up to 30 keV, manifested as a softening with increasing energy so that the spectral shape could be described by either a broken power-law or a log-parabolic fit with equal statistical goodness of fit \citep{2008ApJ...682..775R}. \\
\\
A search for multi-band optical flux and colour variations from PG 1553+113 on IDV timescales was performed and no such variations were noticed  \citep{2012MNRAS.425.3002G}. Extensive multi-band optical observations of PG 1553+113 were done during July 2013 -- August 2014 using five telescopes in Asia and Europe. No genuine IDV in flux or colour was detected in any of these observations whereas clear LTV in flux and colour were detected in all bands in which observations were conducted (BVRI)  \citep{2016MNRAS.458.1127G}.
Recently multi-band optical flux and spectral variations of this BL Lac object were examined using observations carried out on 8 nights in April 2016 and only in one night IDV was detected in V and R band fluxes when a mean optical spectral index $\alpha_{optical} \sim$ 0.83$\pm$0.02 with a maximum variation of 0.21 was found \citep{2019ApJ...871..192P}.
PG 1553+113 has a radio emitting jet extending at least 20 pc to the northeast  \citep{2003AJ....125.1060R}. In VLBA interferometric observations at 22 GHz, the source showed a quiet compact-jet structure  \citep{2012arXiv1205.2399T,2014ApJ...797...25P}.
A marginal quasi-periodic oscillation (QPO) in PG 1553+113, of central period $\sim 2.2$ years in the observer's frame (1.48 years in the source frame) was suggested to be present in the $\gamma-$ray flux with observations done by {\it Fermi}-LAT \citep{2015ApJ...813L..41A}. This possible QPO was supported  by the presence of  similar patterns in the radio and optical regimes \citep{2015ApJ...813L..41A}, leading to even more attention to this source. 
A model invoking a binary system of SMBHs can explain this type of quasi-periodicity \citep{2017ApJ...851L..39C}. This campaign further continues in multiple wavelengths, but only five main peaks have been observed over nine years.  Other possible origins of these peaks in the light curve are a helical jet model and a two jet model \citep{2018ApJ...854...11T}.  \\
\\
We have been running an extensive X-ray IDV study of blazars for over a  decade, using timing data from various X-ray telescopes, e.g. {\it XMM-Newton, Chandra, NuSTAR, Suzaku} \citep{2010ApJ...718..279G,2015MNRAS.451.1356K,2016NewA...44...21B,2016MNRAS.462.1508G,2017ApJ...841..123P,2018ApJ...859...49P,2018MNRAS.480.4873A,2019ApJ...884..125Z}.     In blazars, and particularly in BL Lacs, the radiative emission is dominated by the Doppler boosted relativistic jet rather than the accretion disc. Unlike for  radio IDV, where extrinsic interstellar scintillation plays a role \citep[e.g.][]{1992A&A...258..279Q}, IDV in X-ray bands for blazars is an intrinsic  phenomenon which may be explained by changes in the density of relativistic particles or irregularities in the magnetic field resulting from interaction with shocks and/or turbulent jets \citep{2014ApJ...780...87M,2015JApA...36..255C}. In general, LTV in blazars can be explained by the shock-in-jet model \citep{1985ApJ...298..114M} along with geometric effects such as an  overall bending of the jet. Under this project, we present here X-ray IDV, STV and multi-wavelength LTV studies of one the most observed TeV blazars, PG 1553+113, with observations taken by {\it XMM-Newton}. Although the above-mentioned and other other possible explanations for IDV for blazars in the X-ray band have been posited, there is no consensus on the physics underlying these fast variations.  By gathering all of the data taken with {\it XMM-Newton}, most of which has not been examined for the presence of IDV, and subjecting it to a new and uniform analysis we are able to check whether or not this source actually frequently shows significant rapid X-ray variations.  Further, we examine the power-spectra of those variations, when present.  These analyses shed light on the physics of this emission, as discussed in Section 5.  \\
\\
The paper is structured as follows. Section 2 discusses the data selection from the  {\it XMM-Newton} archive  and the data reduction; in Section 3 we briefly describe the analysis techniques used. Results and discussion are provided in Section 4 and Section 5, respectively.

\begin{table*}
\centering
\caption{Observation log of \textit{XMM-Newton} X-ray data for PG 1553+113}
 \label{tab:obs_log}
 \begin{tabular}{cccccrcl}
\hline
Observation & Observation &  Filter & Window & Pileup & GTI & Bin Size & OM Filter \\
    ID      &  date       &         &  mode  &        & (ks)&    (s)   &  \\\hline
 0656990101& 2010.08.06& Thin1  &Small & No  &21.4   &200 & 1,2,3,4,5,6 \\
 0727780101&2013.07.24 & Medium &Small & No  &32.9   &200 & 1,2,3,4,5,6 \\
 0727780201&2014.07.28 & Medium &Small & No  &34.7   &200 & 1,2,3,4,5,6 \\
 0761100101&2015.07.29 & Thin1  &Full  & Yes &130.3  &500 & 1,2,3,4,5,6 \\
 0761100201&2015.08.02 & Thin1  &Full  & Yes &123.0  &500 & 1,2,3,4,5,6 \\
 0761100301&2015.08.04 & Thin1  &Full  & Yes &135.8  &500 & 1,2,3,4,5,6 \\
 0761100401&2015.08.08 & Thin1  &Full  & Yes &133.5  &500 & 1,2,3,4,5,6 \\
 0761100701&2015.08.16 & Thin1  &Small & Yes &88.4   &300 & 1,2,3,4,5,6 \\
 0761101001&2015.08.30 & Thin1  &Full  & Yes &127.0  &500 & 1,2,3,4,5,6 \\
 0727780301&2015.09.04 & Medium &Small & Yes & 28.4  &200 & 1,2,3,4,5,6 \\
 0727780401&2016.08.17 & Medium &Small & Yes & 28.4  &200 & 1,2,3,4,5,6 \\
 0790380501&2017.02.01 & Thin1  &Full  & Yes & 56.8  &300 & 1,2,3,4,5,6 \\
 0790380601&2017.02.05 & Thin1  &Full  & Yes & 132.3 &500 & 1,2,3,4,5,6 \\
 0790380801&2017.02.07 & Thin1  &Full  & Yes & 134.3 &500 & 1,2,3,4,5,6 \\
 0790380901&2017.02.11 & Thin1  &Full  & Yes & 135.2 &500 & 1,2,3,4,5,6 \\
 0790381401&2017.02.13 & Thin1  &Full  & Yes & 138.4 &500 & 1,2,3,4,5,6 \\
 0790381501&2017.02.15 & Thin1  &Full  & Yes & 138.9 &500 & 1,2,3,4,5,6 \\
 0790381001&2017.02.21 & Thin1  &Small & Yes & 95.5  &500 & 1,2,3,4,5   \\
 0727780501&2017.08.22 & Medium &Small & Yes & 28.5  &200 & 1,2,3,4,5,6 \\
 0810830101&2018.08.25 & Medium &Small & Yes & 33.2  &200 & 1,2,3,4,5,6 \\
\hline
\end{tabular}
\end{table*}

\section{\textit{XMM-NEWTON} ARCHIVAL DATA SELECTION AND REDUCTION} \label{sec:data}

\subsection{Data Selection Parameters} \label{subsec:dataselect}
\noindent
We selected the extreme TeV blazar PG 1553+113 from the TeV source catalogue (TeVCat\footnote{http://tevcat.uchicago.edu/}). The source was observed by {\it XMM-Newton} on many occasions continuously for extended periods of time which is extremely useful for our study.
The {\it XMM-Newton} satellite is capable of doing simultaneous observation in multiple energy bands
and also has a large field of view (FOV).
We took data from the European Photon Imaging Camera (EPIC) PN detector in the energy range (0.2$-$15 keV) and the Optical Monitor (OM) in the wavelength range (170nm$-$650nm). EPIC has three coaligned X-ray CCD-based instruments, EPIC-PN and two EPIC-MOS covering a  30 arcmin FOV \citep{2001A&A...365L..18S}. The OM has three ultraviolet (UVW1, UVM2, UVW2) and three optical (V, U, B) filters with a 17 arcmin FOV \citep{2001A&A...365L..36M}, which  provide simultaneous observations that allow us to test for correlations between X-ray, optical, and UV bands. We considered only the EPIC-PN data for our X-ray study because it is more sensitive and efficient than the EPIC-MOS detector \citep{2001A&A...365L..18S}. In total, {\it XMM-Newton} made 22 observations of the source, taken between August 2010 -- August 2018, and we downloaded the data of all these observation IDs from the HEASARC (High Energy Astrophysics Science Archive Research Centre) Data Archive. \\
\\
We excluded those observations which have either poor image quality or a good time interval (GTI) of less than 20 ks. Using these selection criteria,  we still had 20 extended observations for our study, in which the GTI ranges from 21.4 ks to 138.9 ks, binned into intervals of 200--500 s. The detailed observation log provided in Table 1 shows the total observing time was more than 450 h. For all of these 20 observation IDs, the OM observations were performed in all  six filters, except for observation ID 0790381001 for which no data was available from the UVW2 filter. In every OM observation, each filter has two sets of exposure modes. One mode is imaging, in which the exposure time is 5000 seconds. The other mode is the fast one which has exposure times in the range of 3400--4400 seconds.
The detailed observation log is provided in Table 1.

\subsection{Data Reduction} \label{subsec:datareduction}
\noindent
We have used the standard procedure of the {\it XMM-Newton} Science Analysis System (SAS) version 16.1.0 to reprocess the Observation Data Files (ODF) and to calibrate the summary file with the updated Calibration Current File (CCF)\footnote{http://www.cosmos.esa.int/web/xmm-newton/sas-threads}. We used the {\it epchain} pipeline to generate the event file of the source. To check for high background soft-flares, we extracted light curves in the high energy range 10$-$12 keV taking the full-frame of the CCD and found flaring in some of the observation IDs. Next, we generated GTI files where the count rate was less than 0.4 ct sec$^{-1}$, and these files contain information on free soft proton
flares. We used  the GTI files information and event files as input, and thus produced clean event files which are  used to generate the X-ray light curves. \\
\\
We used standard imaging mode data from the OM for our UV and optical studies. For the OM data reduction we used the standard SAS routine {\it omichain}, which is a {\it Perl} script. After reduction, we got a list of all objects which are present in the OM FOV. We then located our source and the corresponding source count rate and equivalent instrumental magnitude for further use.

\subsection{Light Curve Generation} \label{subsec:LC}
\noindent
With the help of the clean event files, we generated a full-frame image of the CCD and selected a circular region of radii 30--45 arcsec centered around the source. These circular regions contain most of the point spread function (PSF). Since this blazar is a  bright source, there is a real possibility of pileup, where there is a deficit of single events and an excess of double events. We therefore examined the pileup effect for each observation by using the {\it epatplot} SAS task and found pileup mainly in those observation IDs which have longer exposures. For pileup correction we follow the standard procedure and consider an annular region that excludes the over-exposed central region in any observation ID \footnote{https://www.cosmos.esa.int/web/xmm-newton/sas-thread-epatplot}. So for those cases we use a central annulus with inner radius range of 7--12 arcsec instead of a circular region; the annular size that minimizes pileup for each observation is determined iteratively.  A background region is selected within the same CCD chip on which the source is present and is taken to be a blank region of the same radius as  the source and less affected by the source region. To get the source count we subtract this background noise level from the nominal measurement, which includes those extra counts. Using the {\it eveselect} SAS task we applied the condition (PATTERN ${<}$ 4) in the energy range 0.3--10 keV. Finally, we use {\it epiclccorr} task and obtained the corrected event file. From this file we generated light curves (LCs). We made LCs in three different energy bands: 0.3--2.0 keV (soft X-ray band), 2.0--10 keV (hard X-ray band), and 0.3--0 keV (total X-ray band).  A typical LC is shown in Fig.\ 1 and the other 19 are in Supplemental figures.

\section{ANALYSIS TECHNIQUES} \label{sec:AnaTech}

\noindent
In this section we introduce various analysis methods and apply those to the data. The results obtained by these methods are discussed in the following Section 4.

\subsection{Excess Variance} \label{subsec:EVar}
\noindent
Blazars show rapid and strong flux variability over the whole electromagnetic (EM) spectrum on different time scales. Finite uncertainties $\sigma_{err, j}$ are produced in all $N$ individual measured flux values of $x_{j}$ in a LC due to the measurement error in the observations. An additional variance generated in the individual flux measurements, and due to these uncertainties, leads to the calculation of an excess variance $\sigma_{XS}^2$ which is the measure of the strength of variability in the source flux, and so can be considered to be the intrinsic source variance \citep{2002ApJ...568..610E, 2003MNRAS.345.1271V}. \\

\noindent
The fractional rms variability amplitude, $F_{var}$, which is the square root of the normalised excess variance, \citep{2003MNRAS.345.1271V} is given by,
\begin{equation}
F_{var} = \sqrt{\frac{S^2 - \overline{\sigma_{err}^2}}{{\bar{x}^2}}} ,
\end{equation}
in terms of the sample variance, $S^2$, the mean square error, $\overline{\sigma_{err}^2}$, and the mean of the measurements, $\bar{x}$.
Here we have used the expression of uncertainty in $F_{var}$ from \citet{2003MNRAS.345.1271V}

\begin{equation}
(F_{var})_{err}=\sqrt{ \left( \sqrt{\frac{1}{2n}} \frac{ \overline{\sigma_{err}^{2}}
}{\bar{x}^{2}F_{var} } \right)^{2}+\left( \sqrt{\frac{\overline{\sigma_{err}^{2}}}{n}}\frac{1}{\bar{x}}\right)^{2}} ,
\end{equation}
for $n$ measurements.

\subsection{Variability Timescale} 
\noindent
To calculate the shortest flux variability time, we have used halving/doubling timescales from
\begin{equation}
F(t_1)=F(t_2) 2^{(t_1-t_2)/\tau} ,
\end{equation}
\noindent
where $\tau$ is a characteristic halving/doubling timescale and $F(t_{1})$ and $F(t_{2})$ are the fluxes of the LC at times $t_{1}$ and $t_{2}$, respectively. Here we only consider changes where the differences in flux at $t_{1}$ and $t_{2}$ is greater than 3 $\sigma$  \citep{2011A&A...530A..77F}. From the value of $\tau$, we can estimate the size of X-ray emitting region.

\subsection{Discrete Correlation Functions} \label{sec:DCF}
\noindent
The Discrete Correlation Function (DCF) analysis is used to find possible time-lags and cross-correlations between LCs of different energy bands. This technique was introduced by \citet{1988ApJ...333..646E} and later modified by \citet{1992ApJ...386..473H} to find better error estimates. In general, astronomical LCs are unevenly binned, and, for such LCs this technique is very useful as it can be used for unevenly sampled data.
Details about the computation of the DCF we employ here are provided in \citet{2017ApJ...841..123P}. \\
\\
A DCF value ${>}$ 0 indicates that there is some correlation between two data trains, with a value of $+1$ indicating a perfect correlation.   A DCF value ${<}$ 0 means that the data trains are anti-correlated, while if the DCF = 0  there is no correlation between them. When using the same data trains one has an auto-correlation function ACF with a guaranteed peak at $\tau$ = 0 (or no time-lag) but other strong peaks in an ACF indicate a possible periodicity.  \citet{2014MNRAS.441.1899F} and \citep{2014MNRAS.445..428M} demonstrated that DCF values should be considered critically and interpreted carefully, especially when the significance of the result is not estimated. Uncertainties in the data samples were estimated by a model independent Monte Carlo method \citep{1998PASP..110..660P}.

\subsection{Hardness Ratio}
\label{sec:HR}
\noindent
A hardness ratio (HR) analysis provides a model-independent method for learning about spectral variations of the source. We divided the X-ray light curves into two energy bands: a soft one (0.3$-$2.0 keV) and a hard band (2.0$-$10 keV), and then calculated  the hardness ratio (HR) as
\begin{equation}
HR = \frac{(H-S)}{(H+S)} .
\end{equation}
Here S and H are the net counts $s^{-1}$ in the soft and hard energy bands, respectively. \\
\\
We used a standard  $\chi^2$ test for testing for spectral variations in the HR, evaluating,
\begin{equation}
\chi^2 = \sum_{k=1}^{n} \frac{(x_k - \bar{x})^2}{\sigma_k^2}.
\end{equation}
Here $\bar{x}$ is the mean value of HR, ${x_k}$ is the HR value for the $k^{th}$ data point
, and ${\sigma_k}$ is the corresponding error.

\subsection{Duty Cycle}
\noindent
The duty cycle (DC)  provides a direct estimation of the fraction of time for which a source has shown variability. We have estimated the DC of PG 1553+113 by using the standard approach \citep{1999A&AS..135..477R}. For these DC calculations, we considered only those LCs which were continuously monitored for at least about 6 hours, with
\begin{equation} 
DC = 100\frac{\sum_\mathbf{i=1}^\mathbf{n} N_i(1/\Delta t_i)}{\sum_\mathbf{i=1}^\mathbf{n}(1/\Delta t_i)}  {\rm per~cent} 
\end{equation}
here $\Delta t_i = \Delta t_{i,obs}(1+z)^{-1}$ is the redshift corrected GTI of the source observed having the $i^{th}$ Observation ID, and $N_i$ takes the value 1 whenever IDV is detected, and 0 when it is not; when the fractional variability amplitude , $F_{var}$ is greater than 3 times $(F_{var})_{err}$, then we consider the LC to show geniune IDV. Computation of the DC has been weighted by the GTI $\Delta t_i$ for the $i^{th}$ observation ID, as the GTI is different for each observation.

\subsection{Power Spectral Density}
\noindent
The power spectral density (PSD) shows the distribution of the variability power   as a function of temporal frequency. The PSD is a form of periodogram analysis of LCs that allows
 for the  characterization of the temporal variations in flux, including any exact periodicity or quasi-periodic oscillations (QPOs). In this method one  first calculates the square of the Fourier transform of the given LC and then  fits the red-noise variability of the PSD to a power-law. A significant QPO may be present whenever a peak in the PSD rises at least  3$\sigma$ (99.73\%) above the red-noise level of the PSD.  To obtain the  PSDs of these LCs we follow the procedure given by \citet{2005A&A...431..391V} as implemented in \citet{2019ApJ...884..125Z}. We measured the Poisson (white) noise level of the power spectrum in the periodogram which causes the observed spectrum to flatten at high frequencies as the power in the red noise spectrum of the source becomes comparable to the power in the flat Poisson noise spectrum. Hence, we fit only the frequency range below which red noise dominates over the  Poisson noise level.
Using this recipe we found the PSDs; the unit of the periodogram is $ \rm{(rms/mean)^{2}(Hz)^{-1}}$. In most variable AGNs, their PSDs can be characterized by a single power-law (although some are better fit by bending power-laws). We assume the former, and search for a relationship between the temporal frequency, $\nu$, and the red-noise part of the PSD P($\nu$) of   the power-law form P($\nu$) = N$\nu^{\alpha}$, where $N$ is the normalisation and $\alpha$ is the power-law spectral index ($\alpha$ $\leq$ 0) \citep{1989ARA&A..27..517V}, or the slope of the fit to logarithms of $P$ and $\nu$. 

\label{fig:fig1c}
\begin{figure}
\centering
\includegraphics[width=8cm, height=8cm]{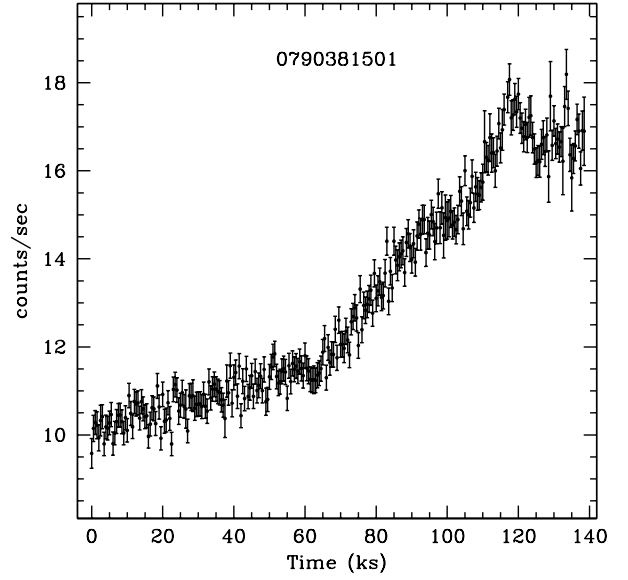}
\caption{\label{1_a} A sample \textit{XMM-Newton}  light curve of the TeV blazar PG 1553+113, labelled with its observation ID.  The light curves for all observations appear in online supplemental material.}
\end{figure}

\begin{table*}
\centering
\caption{X-Ray Variability and Timescale Parameters}

 \label{tab:obs_log}
 \begin{tabular}{cccccccc}
\hline
Observation ID & \multicolumn{3}{c}{F$_{var}${(\%)}} & Significance$^a$ & Variable & $|\tau|^b$(ks) & $|\tau|_{cor}^c$(ks)
\\\cline{2-4}
               &  Soft & Hard & Total &  \\
               &  (0.3-2 keV)& (2-10 keV)& (0.3-10 keV) & \\\hline
0656990101 &~0.82$\pm$0.22 &~2.90$\pm$0.89  &~1.06$\pm$0.22 & 4.81  &V  & 11.37$\pm$~3.27 & ~7.63$\pm$2.40   \\
0727780101 & ....$^d$      &~1.33$\pm$0.37  & ....          &    	  & 	  & 23.41$\pm$~7.39 & 15.71$\pm$5.38   \\
0727780201 & ....          & ....           &~0.68$\pm$0.16 & 4.25  &V  & 13.02$\pm$~3.19 & ~8.74$\pm$2.38   \\
0761100101 &~2.04$\pm$0.17 & ....           &~2.07$\pm$0.16 & 12.93 &V  & ~2.76$\pm$~0.79 & ~1.85$\pm$0.58   \\
0761100201 &~0.86$\pm$0.19 & ....           &~0.46$\pm$0.19 & 2.42  &NV & ~5.64$\pm$~1.88 & ~3.78$\pm$1.36   \\
0761100301 &~2.40$\pm$0.18 &~0.59$\pm$0.59  &~2.24$\pm$0.17 & 13.18 &V  & ~2.86$\pm$~0.91 & ~1.92$\pm$0.66   \\
0761100401 &~0.90$\pm$0.19 &~1.69$\pm$0.66  &~0.25$\pm$0.19 & 1.32  &NV & 14.11$\pm$~3.55 & ~9.47$\pm$2.64   \\
0761100701 &~2.89$\pm$0.16 &~3.18$\pm$0.55  &~2.84$\pm$0.16 & 11.72 &V  & ~3.63$\pm$~1.16 & ~2.43$\pm$0.84   \\
0761101001 &~4.82$\pm$0.19 &~1.49$\pm$0.67  &~4.75$\pm$0.18 & 26.39 &V  & ~2.54$\pm$~0.65 & ~1.70$\pm$0.48   \\
0727780301 &~2.76$\pm$0.29 & ....           &~2.58$\pm$0.29 & 8.91  &V  & 22.04$\pm$~6.17 & 14.79$\pm$4.54   \\
0727780401 &~1.15$\pm$0.27 & ....           &~0.73$\pm$0.26 & 2.81  &NV & 33.31$\pm$10.58 & 22.35$\pm$7.69   \\
0790380501 &~3.15$\pm$0.43 &~4.27$\pm$1.63  &~2.84$\pm$0.42 & 6.72  &V  & ~5.64$\pm$~1.72 & ~3.78$\pm$1.26   \\
0790380601 &~3.55$\pm$0.21 &~6.83$\pm$0.81  &~3.46$\pm$0.21 & 16.48 &V  & ~4.09$\pm$~1.28 & ~2.75$\pm$1.22   \\
0790380801 &17.87$\pm$0.23 &26.22$\pm$0.85  &18.47$\pm$0.22 & 83.96 &V  & ~2.39$\pm$~0.69 & ~1.61$\pm$0.50   \\
0790380901 &~2.55$\pm$0.22 &~4.97$\pm$0.82  &~2.66$\pm$0.21 & 12.67 &V  & ~5.55$\pm$~1.80 & ~3.73$\pm$1.31   \\
0790381401 &~1.21$\pm$0.25 &~6.68$\pm$0.98  &~1.74$\pm$0.24 & 7.25  &V  & ~6.07$\pm$~1.85 & ~4.08$\pm$1.35   \\
0790381501 &17.65$\pm$0.16 &30.75$\pm$0.54  &18.72$\pm$0.15 & 124.8 &V  & ~5.37$\pm$~1.54 & ~3.60$\pm$1.13   \\
0790381001 &~5.36$\pm$0.18 &~4.76$\pm$0.66  &~5.31$\pm$0.17 & 31.24 &V  & 10.14$\pm$~3.31 & ~6.81$\pm$2.40   \\
0727780501 &~0.72$\pm$0.13 &~1.48$\pm$0.39  &~0.63$\pm$0.12 & 5.25  &V  & ~6.32$\pm$~1.98 & ~4.24$\pm$1.45   \\
0810830101 &~1.04$\pm$0.18 & ....           &~1.08$\pm$0.17 & 6.35  &V  & 10.06$\pm$~2.79 & ~6.75$\pm$2.05   \\
\hline
 \end{tabular}\\
 Note: V= Variable  NV= Non-variable\\
$^a$ Significance = $F_{var}/  (F_{var})_{err}$ for the total band   \\
$^b |\tau|$ = Observed halving/doubling time scale in ks\\
$^c |\tau|_{cor}$ = Redshift corrected halving/doubling time scale: $|\tau|_{cor}$ = $|\tau|$/(1 + z) \\
$^d$ .... indicates that the variance arising from measurement error is greater than the total variance of the LC\\

\end{table*}

\label{fig:fig2}
\begin{figure*}
\centering
\includegraphics[width=18cm, height=3cm]{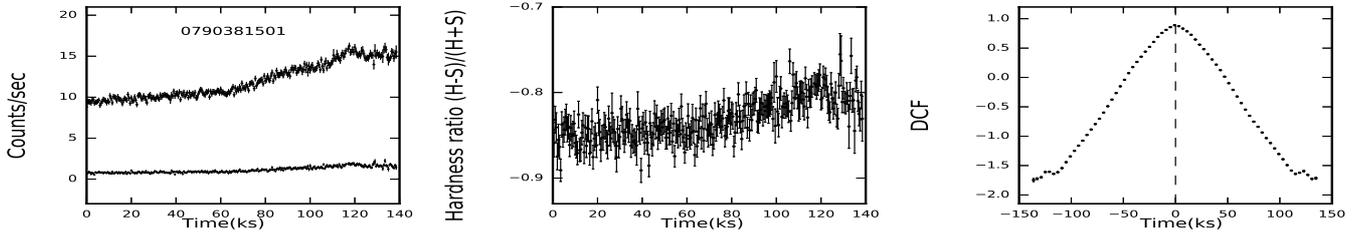}
\caption{\label{2_a}Soft energy (0.3$-$2 keV, upper plot, denoted by open circle) and hard energy (2$-$ 10 keV, lower plot, denoted by asterisk) light curves for the TeV blazar PG 1553+113 (left panel) for a sample XMM-Newton observation ID 0790381501; corresponding hardness ratio (middle panel); discrete correlation function between those bands (right panel). Similar plots for all the observations appear in online supplemental material.}
\end{figure*}

\section{Results}
\noindent
We now employ the various analysis methods discussed in Section 3 to the data described in Table 1 to find the following results.  

\subsection{Intraday X-ray Variability}
\noindent
In this section we discuss the results of our studies of the flux, spectral and cross-correlated X-ray variabilities of these 20 {\it XMM-Newton} extended observations of the source.

\subsubsection{Flux Variability}
\noindent
We generated X-ray LCs for all the selected 20 observations: an exemplary LC of observation ID 0790381501 is shown in Fig.\ 1 and the remaining 19 LCs are displayed in online supplemental material. On visual inspection, only a few LCs show clear evidence of variability on these IDV timescales. To more fairly interrogate the variability on IDV timescales and to quantify their amplitudes, we have used the excess variance method briefly described in section 3.2, and the results obtained are reported in Table 2. We consider a LC variable only when the sample variance is more than the mean square error and the fractional amplitude variability, $F_{var}$, is greater than 3 times $(F_{var})_{err}$, following \citet{2015ApJ...811..143P}.  With this definition we see that  16 out of 20 LCs show IDV (based on the entire X-ray band). \\
\\
For further analysis, as mentioned above, we divided the whole X-ray energy range (0.3$-$10 keV) into two energy bands: soft  (0.3$-$2 keV) and hard  (2$-$10 keV)
and plot them in the left panels of Figure 2, which shows an example of these LCs. 
We calculated $F_{var}$ values and their errors in the hard and soft bands separately, as well as in the entire X-ray energy range, with results reported in Table 2. 
In the soft band, all of the 18 (out of 20) observations for which the sample variance is more than the mean square error of the observation ID show significant variability, as they also have $F_{var} > 3(F_{var})_{err}$.  In the hard band the count rates are lower and only 14 of the 20 have sample variances exceeding the mean square error; of those, 10 also satisfy the criterion $F_{var} > 3(F_{var})_{err}$.
  However, when both bands show variations, the hard X-ray band variability is generally greater than  the corresponding soft band (10 out of 13 cases). Two observations (IDs 0790380801 and 0790381501) show substantial large amplitude IDV, with $F_{var} >  $18 \%.\\
\\
In our analysis of the blazar PG 1553+113, we found the shortest X-ray flux observed doubling time to be 2.4$\pm$0.7~ks for the observation ID 0790380801. During three other observations the doubling times are comparably short ($< 3$ ks) and most of them are under 6.5~ks but in one of the observations (ID  0727780401) any doubling time is larger than the data length.   These short doubling times are a key result and throughout the rest of the paper we will use the shortest doubling time of 2.4$\pm$0.7~ks to estimate the various physical parameters related to this emission from the blazar.

\subsubsection{Cross-correlated Variability}
\noindent
We have used the DCF technique to determine cross correlations and thus search for any time lags between the soft and hard X-ray energy bands for each of the observations of PG 1553+113. The DCF plot for the example LC is shown in the right panel of Figure 2 and in this case there is a strong correlation at 0 lag.  However, presumably because the count rates in the hard band are usually so low and do not show obvious variations, only one more of the DCFs plotted in the supplemental figures yields a value above 0.5 at 0 lag, so in general, no correlations can be claimed.  Still, for none of the LCs were there any significant DCF values at any non-zero time lags between the hard and soft bands.  This result is consistent with the hypothesis that emission in both energy bands occurred at the same region and arose from the same population of leptons \citep{2017ApJ...841..123P}, but by no means confirms it.

\subsubsection{Spectral Variability}
\noindent
As a simple check for X-ray spectral variation on IDV timescales we looked for changes in the HR using a standard $\chi^{2}$ test, as discussed in section 3.4. We considered spectral variation to be present for those observations in which $\chi^2$ ${>}$ $\chi^2_{99,\nu},$ where $\nu$ is number of DOFs (degrees of freedom). The HRs as functions of time for individual observations are plotted for the example LCs in the middle panel of Figure 2. Three of these observations, all of which have large amplitude variability in fluxes, showed significant variations in HR with time;  in all of those cases the spectra gets harder with an increase in brightness.  However, the majority of the observations did not show any significant HR variations, which is not surprising for situations where there are no large amplitude flux variations.

\subsubsection{Duty Cycle}
\noindent
Our method of estimation of the DC is mentioned above. Based on the $F_{var}$ test, we took the value of $N_i$ as either 1 if variable, or 0 if not. In the whole X-ray energy band (0.3--10 keV) 16 out of 19 observations showed variability (as one LC had measurement errors exceeding the nominal variation). Using this  approach we found the X-ray DC of this source to be  $\sim$84\%, illustrating that detectable IDV over this long time duration (2010--2018) was common.  In an earlier analysis of a subset of 6 of these LCs  \citet{2017MNRAS.466.3762R} only one was noted to show IDV, but with our approach two others would be considered as variable.

\subsection{Power Spectral Density}
\noindent
To characterize the type of noise present  in the variations on intraday timescales and to attempt to search for any QPOs over those spans, we performed  PSD analyses of all  20 X-ray observations studied here. From the PSD plots (an example is shown in Fig.\ 3), we found all observations to have some degree of red-noise dominance and there is no sign of quasi-periodicity in any observation.\\
\\
The slopes ($\alpha$) of the PSDs are in the range of $-0.14 \pm 0.18$ (essentially white noise) to $-2.36 \pm 0.25$ steep, which is commonly seen in AGN X-ray variability \citep{2012A&A...544A..80G}, and in some observations its value is $\sim -$1 (close to flicker noise).  The values of the power-law slopes and normalisation constants for the all 20 observations are reported in Table 3. The normalisation constants are in the range of $-9.24 \pm 0.96$  to $-0.12 \pm 0.51$ for all observations, and the mean value of the PSD  spectral index for PG 1553+113 is $-1.21 \pm 0.26$.

\subsection{Short and Long Term Variability}
\noindent
In this section we discuss  the flux and spectral variabilities on STV and LTV timescales  by using all 20 {\it XMM-Newton} observations of the blazar PG 1553+113.

\begin{figure}
\centering
\includegraphics[width=8cm, height=8cm]{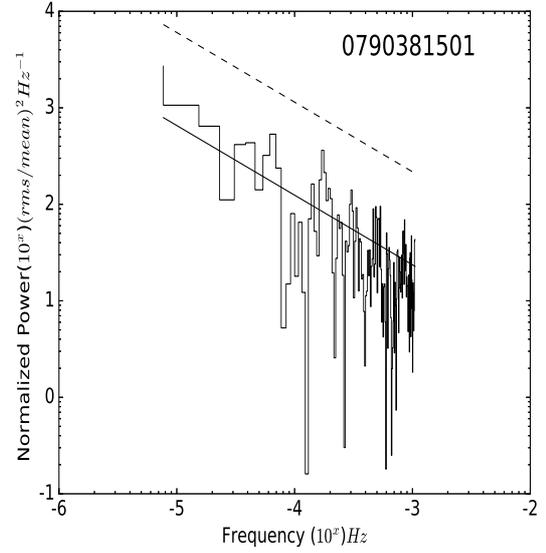}
\caption{\label{5} Power spectral density of the example X-ray light curve in the full energy range (0.3--10 keV), with the continuous line showing the fitted red-noise slope to the power spectral density and the dashed line showing a 3 $\sigma$ level above that noise; any measurements above that would provide a hint of a quasi periodic oscillation.  These plots for all the observations appear in online supplemental material.}
\end{figure}

\begin{figure}
\centering
\includegraphics[width=8cm, height=8cm]{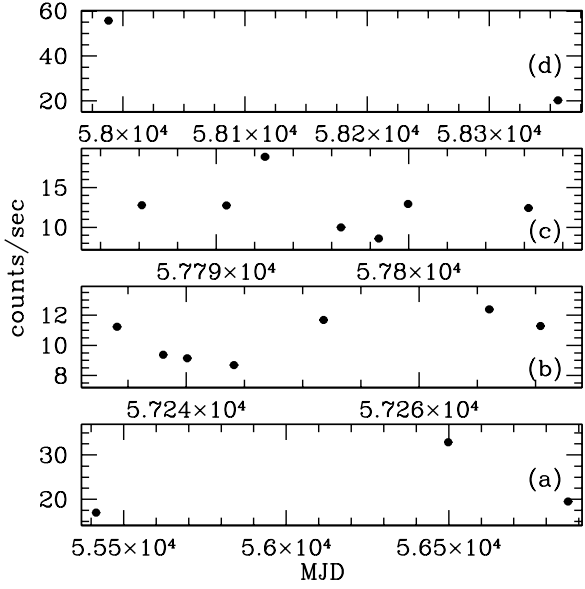}
\caption{\label{6} Short term variability plots in 0.3--10 keV. The observation spans are: (a) 6 August 2010 to 28 July 2014; (b) 29 July 2015 to 4 September 2015; (c) 1 -- 21 February 2017;  (d)  observations taken on 22 August 2017 and 25 August 2018. Errors are not visible as they are smaller than the symbol size.}
\end{figure}

\begin{figure}
\centering
\includegraphics[width=8cm, height=8cm]{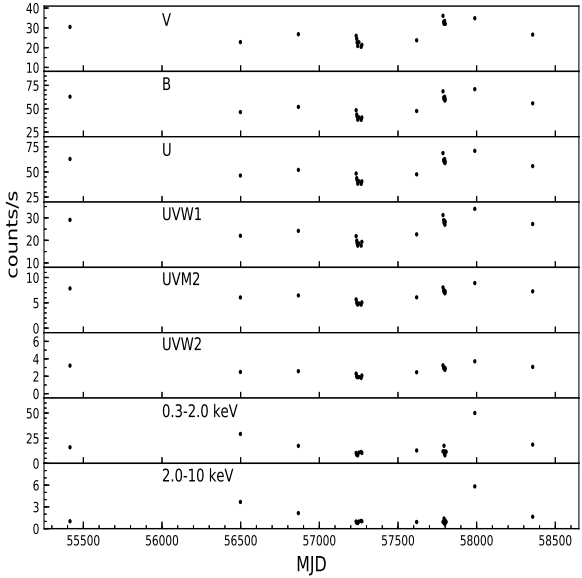}
\caption{Optical/UV and X-ray long term variability (2010--2018). Going from the uppermost panel to the lowermost, the energies rise from the labeled optical bands  through the UV bands to the soft and hard X-ray bands.}
\end{figure}

\begin{figure}
\centering
\includegraphics[width=8cm, height=8cm]{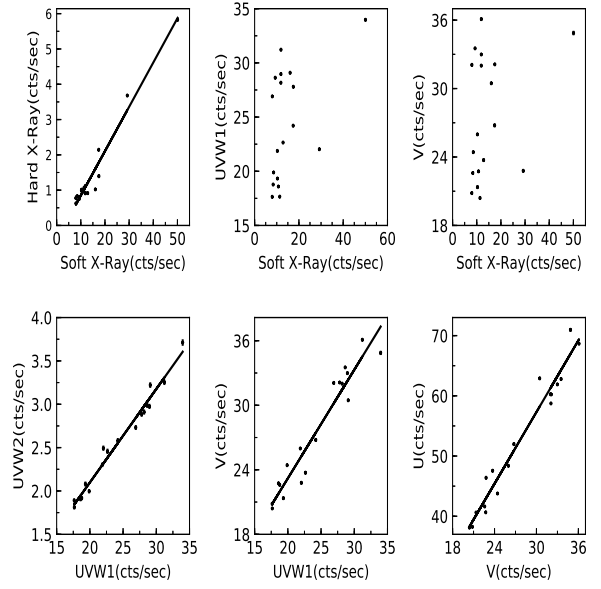}
\caption{Correlations between several  X-ray, optical and UV bands, as labeled on the axes.}
\end{figure}

\subsubsection{Short Term X-ray Flux Variability}
\noindent
We divide the complete set of observations into four different segments to study the X-ray STV nature of PG 1553+113 as shown  in Figure 4. In panel (a), the first three average fluxes are displayed though they cover essentially the first four years.  A denser sampling in panel (b) shows seven observations made over five weeks and in panel (c) seven observations were made in only three weeks. Panel (d) shows the last 2 observations, which were taken essentially a year apart. We omit plotting one data point taken on 17 August 2016 which has an average flux 13.74$\pm$0.34 because it fell between the second and third grouping and had considerably larger gaps with both.  The STV plots in all four panels of Fig.\ 4 show clear evidence of large flux variations on STV timescales. Considering all the measurements, the faintest to strongest total X-ray fluxes differ by a factor of  6.7.

\begin{table}
\centering
\caption{Power-law fits to the PSDs$^{a}$ of all 20 observations}
 \label{tab:region}
 \begin{tabular}{lcc}
  \hline
Obs Ids & ($\alpha$) &  log($N$)$^{b}$ \\

\hline
0656990101 & $-$2.12 $\pm$ 0.16  & $-$8.54 $\pm$ 0.65\\
0727780101 & $-$1.13 $\pm$ 0.54  & $-$5.54 $\pm$ 2.21\\
0727780201 & $-$1.26 $\pm$ 0.54  & $-$5.94 $\pm$ 2.09\\
0761100101 & $-$1.22 $\pm$ 0.12  & $-$2.99 $\pm$ 0.42\\
0761100201 & $-$0.67 $\pm$ 0.11  & $-$1.49 $\pm$ 0.38\\
0761100301 & $-$0.81 $\pm$ 0.16  & $-$1.51 $\pm$ 0.59\\
0761100401 & $-$1.39 $\pm$ 0.10  & $-$3.52 $\pm$ 0.38\\
0761100701 & $-$1.91 $\pm$ 0.28  & $-$7.77 $\pm$ 1.18\\
0761101001 & $-$1.24 $\pm$ 0.11  & $-$2.95 $\pm$ 0.38\\
0727780301 & $-$2.36 $\pm$ 0.25  & $-$9.24 $\pm$ 0.96\\
0727780401 & $-$1.57 $\pm$ 0.45  & $-$6.33 $\pm$ 1.75\\
0790380501 & $-$0.93 $\pm$ 0.26  & $-$1.06 $\pm$ 0.89\\
0790380601 & $-$0.91 $\pm$ 0.15  & $-$1.09 $\pm$ 0.53\\
0790380801 & $-$1.21 $\pm$ 0.13  & $-$2.32 $\pm$ 0.47\\
0790380901 & $-$1.54 $\pm$ 0.25  & $-$3.91 $\pm$ 1.04\\
0790381401 & $-$1.43 $\pm$ 0.22  & $-$3.41 $\pm$ 0.84\\
0790381501 & $-$1.06 $\pm$ 0.17  & $-$2.31 $\pm$ 0.62\\
0790381001 & $-$0.64 $\pm$ 0.08  & $-$1.56 $\pm$ 0.31\\
0727780501 & $-$0.72 $\pm$ 0.26  & $-$3.73 $\pm$ 0.89\\
0810830101 & $-$0.14 $\pm$ 0.18  & $-$0.12 $\pm$ 0.51\\\hline
\end{tabular}

\vspace*{0.1in}
\noindent
$^a$ A power-law model is assumed with P(f) $\propto$ f$^{\alpha}$ for $\alpha <$ 0\\
$^b$ Normalisation Constant, (rms/mean.Hz)$^2$

\end{table}

\subsubsection{Long Term Multi-wavelength Flux Variability}
\noindent
We used all the data to analyse the long-term flux behaviour of the {\it XMM-Newton} satellite observations of PG 1553+113 over the period of 2010 to 2018 in optical and UV bands as well as  in X-rays.   We plot the fluxes in counts s$^{-1}$ versus MJD for all the available bands in Figure 5 and calculated their fractional rms variability amplitudes, as discussed in section 3.1.  Here we have estimated the rms variability amplitude (i.e., $\sigma^{2}/m^{2}$ where $\sigma^{2}$ and $m$ are the variance, corrected for the experimental contribution, and mean of the light curve, respectively) for each light curve \citep{{2016NewA...44...21B}}. In Fig.\ 5, from top to bottom the panels correspond to  increasing photon energies. In the X-ray band comparison, the hard band showed a slightly larger amplitude variation compared
 to the soft band and the variabilities in these energies are 87 per cent, and 66 per cent, respectively. On other hand, all the optical/UV bands show nearly identical flux variations, with those corresponding to V, B, U, UVW1, UVM2, and UVW2  being 19, 20, 20, 20.5, 20, and 21 per cent, respectively. \\
 \\
 We also quantified the correlation between X-ray, optical and UV band variations and an informative subset of those results are plotted in Figure 6. The two X-ray bands are strongly correlated, with coefficient $> 0.98$, and comparisons between the UV and optical bands also show strong correlations, with all coefficients $> 0.96$.  However, there are no significant correlations between the X-ray and UV/optical bands, as shown in Table 4, where we give correlation coefficients and probabilities for the null hypothesis that there is no correlation between them (p-values). 

\begin{figure}
\centering
\includegraphics[width=8.5cm, height=8.5cm]{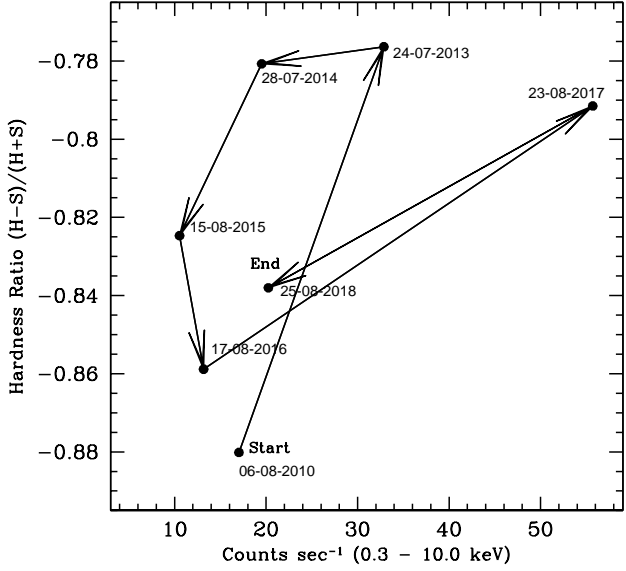}
\caption{\label{6} The hardness ratio plotted against flux at different times.   }
\end{figure}

\subsubsection{Long Term X-ray Spectral Variability}
\noindent
We next consider the variation of the hardness ratio with time.  Seven observations taken from 29 July 2015 to 4 September 2015 have almost the same HR values, so we took an average of these seven HR values for this interval. Similarly, for eight observations taken from 17 August 2016 to 21 February 2017, the HR values are nearly the same and we averaged these eight HR values. By so combining the total of 20 extended observations, we have seven HR versus flux (in 0.3--10 keV) data points, which are shown in Fig.\ 7.

\begin{table}
\centering
\caption{Variability Correlation parameters between  X-ray, Optical and UV  bands}
 \label{tab:region}
 \begin{tabular}{lcc}
  \hline
Bands & Correlation & p-value \\
      & coefficient &  \\
\hline
Soft$-$Hard X-ray & 0.9814 & 0.0\\
Soft X-ray$-$UVW1 & 0.4804 & 0.0436\\
Soft X-ray$-$V & 0.2883 & 0.2459\\
UVW1$-$UVW2 &  0.9887 & 0.0\\
UVW1$-$V & 0.9619 & 0.0\\
V$-$U & 0.9785 & 0.0\\
\hline
\end{tabular}
\end{table}

\section{Discussion and Conclusions}
\noindent
Blazars  often evince strong and rapid flux variations across all EM bands and these changes are useful for understanding the emission mechanisms and estimating the sizes, locations, and sometimes even the structure, of the emitting regions in the jets \citep[e.g.,][]{2003A&A...400..487C}. There are two fundamental classes of theoretical models that can explain intrinsic AGN emission and variability.     Instabilities in or hot spots on accretion discs can explain  IDV and STV in radio quiet quasars and in some  blazars in low flux states, particularly of the FSRQ class \citep[e.g.,][]{1993ApJ...406..420M} but this scenario cannot provide a satisfactory explanation for the rapid, robust variations in flux observed from blazars.    Relativistic jet based models  involving shocks propagating down the jets can explain major LTV \citep{1985ApJ...298..114M, 1995ARA&A..33..163W}.  Weaker fluctuations seen as STV and IDV  can be understood in terms of small wiggles in jet direction or helical perturbations \citep{1992A&A...259..109G, 1992A&A...255...59C, 2017ApJ...841..123P}, jet turbulence \citep{2014ApJ...780...87M, 2015JApA...36..255C}, or a combination of both \citep{2016ApJ...820...12P}. BL Lac objects have a combination of a featureless optical spectrum with the extreme flux variability and their relativistic jet is very close to the line of sight (LOS) of the observer so the strong Doppler boosting can produce major increments in observed flux along with a reduction in the observed variability timescale  \citep{1997ARA&A..35..445U}. \\
\\
To estimate an upper limit to the emission region size $R$, we use the simple causality constraint,
\begin{equation}
R \leq \frac{c ~t_{var} ~\delta}{1+z} ,
\end{equation}
\noindent
where $\delta$ is the Doppler factor.  The
X-ray emission of BL Lacs mostly originates from the jet that is moving at relativistic speeds near the LOS and  values of  $\delta$ for PG 1553+113 have been estimated in previous works.  A variability Doppler factor of 11 was found from  flaring time scales in 15 GHz radio data \citep{2018ApJ...866..137L}, while a larger Doppler factor of between 23 and 35 was estimated through SED fitting with homogeneous one-zone SSC models at different flaring states \citep{2012ApJ...748...46A}.  
Using that range of Doppler factor values,  taking $z = 0.433$ \citep{2010ApJ...720..976D,2019ApJ...884L..31J}, and employing the shortest variability timescale we found ($t_{var}$ = 2.4 ks) in Eqn.\ (7)
above, we obtain that the size of the emission region is in the range of (0.55 -- 1.76) $\times 10^{15}$ cm. \\
\\
The mass of the central SMBH is clearly one of the most fundamental properties of any AGN. In general, spectroscopic techniques involving gas or stellar kinematics and, in particular, reverberation mapping studies, are the most direct  and accurate tools for estimating the mass of a SMBH \citep[e.g.][]{2004ASPC..311...69V}. However in case of BL lac objects such as PG 1553+113, these tools cannot be applied due to the lack of emission lines.  \\
\\
Now it is possible that during an X-ray flare, some fluctuations arise that are related to an orbital period of the accretion disc. Then the shortest orbital period timescale and a plausible lower limit to the size of the emitting region would arise close to the last stable orbit around the SMBH. So one can use the minimum variability timescale to estimate the mass of SMBH \citep{2009ApJ...690..216G}, if the emission arises from the disc, as
\begin{equation}
M / M_{\odot} = 3.23 \times 10^4 \frac{P(s)} {[(r^{3/2} + (a/M)] (1+z)},
\end{equation}
where $a/M$ is the angular momentum of the BH as a dimensionless parameter between $-1$ and $1$, $r$ is the minimum radius of the emitting region in terms of  the gravitational radius, $GM/c^2$, and $P(s)$  is the minimum variability timescale in the flux in seconds.
For a non-rotating (Schwarzschild) SMBH ($a = 0$), the innermost circular orbit is $r = 6$. So for the shortest timescale we found  of  $P = 2.4 \times 10^3$s and with the redshift of PG 1553+113 ($z = 0.433$)  we get an estimate of $\sim3.7 \times 10^6M_{\odot}$ for the SMBH.  However, for the usually assumed maximally rotating accreting Kerr SMBH, $a =0.998$ and $r = 1.2$; then the SMBH mass estimate rises to $\sim2.3 \times 10^7M_{\odot}$  \citep[e.g.,][]{2010ApJ...718..279G}. However, this estimate is not directly applicable to blazars, where the emission mainly arises from the jets that are launched from very close to the SMBH and not from the accretion disc \citep{2008Natur.452..966M}.
But if the variability is due to disturbances that originated in the jet but were injected from the accretion disc near the SMBH \citep[e.g.,][]{2019ApJ...884..125Z}, then in this case Doppler boosting enhances the variability amplitude by $\sim \delta^3$ and decreases the time scale by $\delta^{-1}$ \citep[e.g.,][]{2003ApJ...586L..25G}. So the mass estimation for the SMBH in PG 1553+113 would rise by a factor of {$\delta$}, to $\sim 4 \times 10^7M_{\odot}$, for the Schwarzschild case with the lowest $\delta$ (11), up to $\sim 8 \times 10^8M_{\odot}$ (for the Kerr case and $\delta=35$).
For the following discussion we assume an intermediate value of {$\delta = 25$} and take the upper limit to the size of the emitting region and the mass of SMBH to be {$\sim 1.3 \times 10^{15}$ cm  (from Eq.\ 7) and  $\sim 5.8 \times 10^8M_{\odot}$} (Eq.\ 8, for the Kerr case), respectively. It has been argued that PG 1553+113 is a binary SMBH system, with the primary and secondary SMBH masses of  $\sim(10^8-10^9)M_{\odot}$ and $\sim10^7M_{\odot}$ \citep{2015ApJ...813L..41A, 2018ApJ...854...11T} and so this mass estimate is consistent with the primary in that case. \\
\\
If the fluctuations emerge at a significant distance from the SMBH, then the above approach does not provide clear information about the mass of SMBH. The hard X-ray energy is believed to be generated by synchrotron emission from the HBLs such as PG 1553+113  and some specific parameters can be estimated. Electrons are likely accelerated within the jet by the diffusive shock acceleration mechanism \citep{1987PhR...154....1B} and if this is the case the acceleration timescale in the observer's frame is
\begin{equation}
t_{acc}(\gamma) \simeq 3.78\times10^{-7} \frac{(1+z)}{\delta B} \xi \gamma ~{\rm s},
\end{equation}
where $B$ is the magnetic field, $\gamma$ is the ultrarelativistic electron  Lorentz factor and $\xi$ is the acceleration parameter for the electrons. In the observer's frame, the synchrotron cooling timescale of an individual electron having energy ${\it E = \gamma m_{e}c^{2}}$
\begin{equation}
t_{cool}(\gamma) \simeq 7.74\times10^{8}\frac{(1+z)}{\delta} B^{-2}\gamma^{-1} ~{\rm s}.
\end{equation}

\noindent
The critical synchrotron emission frequency, in the {\it XMM-Newton} satellite energy is in the range \citep{2015ApJ...811..143P} \\
\\
$\nu \simeq 4.2\times10^{6} \delta (1+z)^{-1} B \gamma^{2}$ = $10^{18}\nu_{18}$ Hz, where $0.078< \nu_{18} < 2.42$, and
\begin{equation}
B \geq 0.12~ \nu^{-1/3}_{18} {\rm G} .
\end{equation}
We estimate the value of electron Lorentz factor to be
\begin{equation}
\gamma \leq 3.5\times10^{5} ~\nu^{2/3}_{18} .
\end{equation}
\noindent
The relativistic electrons produce  photons by Compton scattering in Thomson region and the maximum energy would be \citep[e.g.,][]{2018ApJ...859...49P}:
\begin{equation}
E_{T,{\rm max}} \simeq \frac{\delta}{(1+z)} \gamma_{max} m_e c^2 \sim 3~  \nu_{18}^{2/3}  ~{\rm TeV}. 
\end{equation}
\\
We used the model-independent hardness ratio (HR) analysis to investigate the spectral variations in the X-ray range (0.3--10 keV). It is a simple and efficient way to detect  variations in spectra even for relatively low count rates, but does not give any information about physical parameters  that could constrain the reasons for the spectral variation. This study showed no significant change in most of the HR plots, but there were two observations in which we found clear increases in the HR as the counts rate increased.  Thus PG 1553+113 tends to  follow the general  âharder-when-brighter trend  seen in other HSP type blazars \citep{1998ApJ...492L..17P, 2002ApJ...572..762Z, 2003A&A...402..929B, 2017ApJ...841..123P}. \\
\\
We used a DCF analysis technique to search for correlations between soft and hard band X-ray emission and we never found a significant time lag between hard versus soft bands, although the count rates were usually insufficient and the variability too modest to yield significant correlations even at zero lag. In general, the X-rays are well correlated in different energy bands \citep{2018MNRAS.480.4873A}. No time lag has been found in the different energy bands, which is consistent with the emission region being the same for different X-ray energies. \\
\\
The duty cycle for IDV in the {\it XMM-Newton} X-ray energy band (0.3--10 keV) is $\sim$ 84 per cent which showed that the source frequently was at least somewhat variable in the lengthy period spanned by these observations (2010--2018).
The PSDs of all 20 pointed observations X-ray LCs in this energy range are found to be red-noise dominated, but with a range of spectral slopes, and there is no   evidence of a QPO in any of them. Had a QPO signal been detected some interpretation in terms of an origin from a dominant hot-spot on a disc or strong helical features associated with the jet might have been inferred but since QPO signals in AGN LCs are very rare (and usually short-lived) their absence in these LCs does not preclude any such explanations.  The timescales investigated in each observation are too short to constrain any binary black hole model. \\
\\ 
The {\it XMM-Newton} satellite simultaneously conducts observations in the X-ray and optical / UV energy range, helping us to learn some more about the physical process responsible for the emission. This source showed variability in all bands (optical to X-ray) over the long length of these observations. X-ray bands show higher rms variability amplitudes than do the optical and UV bands. We found that the variability amplitude increases with an increase in frequency in X-ray and was essentially constant between the optical/UV bands (with perhaps a slight increase) \citep{2016NewA...44...21B}.  We checked the relations by computing the Pearson correlation coefficients and found  good correlations between the optical and UV bands as well as an independent correlation between the soft and hard X-rays which means that their emission region is similar. But as the X-ray fluxes do not correlate significantly with the optical and UV bands, their emission regions should be different.

\section*{Acknowledgements}
\noindent
We thank the anonymous referee for extensive comments that have led to improvements in the manuscript. This research is based on observations obtained with {\it XMM-Newton}, an ESA science mission with instruments and contributions directly funded by ESA member states and NASA. HG acknowledges  financial support from the Department of Science \& Technology (DST), Government of India, through the INSPIRE faculty award IFA17-PH197 at ARIES, Nainital, India.

\section*{Data Availability}

The data sets were derived from sources in the public domain: [XMM-Newton, https://heasarc.gsfc.nasa.gov/db-perl/W3Browse/w3browse.pl]. The data underlying this article will be shared on reasonable request to the corresponding author.

\section*{Supporting information}

The online version contains the supporting, supplementary data in the form of extensions to Figs. 1--3.

\clearpage

\end{document}